\pdfoutput=1

\documentclass{aastex631}

\accepted{April 1, 2023 }

\graphicspath{{./}{figures/}}

\begin{document}

\title{The Breakthrough Listen Search for Intelligent Life: Nearby Stars' Close Encounters with the Brightest Earth Transmissions}



\author{Reilly Derrick}
\affiliation{Henry Samueli School of Engineering, University of California Los Angeles, Los Angeles, CA 90095, USA}
\affiliation{St. Ignatius College Preparatory, 2001 37th Ave, San Francisco, CA 94116, USA}

\author[0000-0002-0531-1073]{Howard Isaacson}
\affiliation{501 Campbell Hall, University of California at Berkeley, Berkeley, CA 94720, USA}
\affiliation{Centre for Astrophysics, University of Southern Queensland, Toowoomba, QLD, Australia}

\email{reillyderrick1@gmail.com, hisaacson@berkeley.edu}

\begin{abstract}

After having left the heliosphere, Voyager 1 and Voyager 2 continue to travel through interstellar space. The Pioneer 10, Pioneer 11, and New Horizons spacecraft are also on paths to pass the heliopause. These spacecraft have communicated with the Deep Station Network (DSN) radio antennas in order to download scientific data and telemetry data. Outward transmissions from DSN travel to the spacecraft and beyond into interstellar space. These transmissions have encountered and will encounter other stars, introducing the possibility that intelligent life in other solar systems will encounter our terrestrial transmissions. We use the beamwidth of the transmissions between DSN and interstellar spacecraft to perform a search around the past and future positions of each spacecraft obtained from the JPL Horizons System. By performing this search over the Gaia Catalogue of Nearby Stars (GCNS), a catalogue of precisely mapped stars within 100 pc, we determine which stars the transmissions of these spacecraft will encounter. We highlight stars that are in the background of DSN transmissions and calculate the dates of these encounters to determine the time and place for potential intelligent extraterrestrial life to encounter terrestrial transmissions.

\end{abstract}


\section{Introduction} \label{sec:intro}

After launching in 1977, Voyager 1 left the heliosphere in 2012. The Deep Space Network (DSN) continues to send transmissions to the spacecraft, meaning the transmissions traveling from DSN to Voyager 1 continue to travel through interstellar space and are likely to encounter a number of stars in the Milky Way. By determining which stars Voyager 1’s transmissions will encounter, we identify places where possible intelligent extraterrestrial life would encounter terrestrial transmissions and potentially return transmissions toward the Earth. 

Using the JPL Horizons System\footnote{https://ssd.jpl.nasa.gov/horizons/}, we first produce an ephemeris of the spacecraft from Voyager 1’s launch date to its latest available date in the system. Next, we use the beamwidth of the transmissions from DSN to Voyager 1 to perform a cone search around the future and past positions of Voyager 1 in order to determine which background stars its transmissions will encounter. To perform this search we use the three-dimensional positions of over 300,000 stars within 100 pc from the Gaia Catalogue of Nearby Stars (\cite{gaia_gcns}, GCNS). We also ensure that the identified stars will not leave the radius of the search in the time it will take DSN transmissions to reach them.

We perform a similar search over Earth’s four other interstellar spacecraft: Voyager 2, Pioneer 10, Pioneer 11, and New Horizons. Voyager 2 was launched in 1977 and left the heliosphere in 2018. Pioneer 10 launched in 1972 and crossed the orbit of Neptune in 1983 before the project was discontinued in 1997. Pioneer 11 launched in 1973 and crossed the orbit of Neptune in 1990 before the project was discontinued in 1995. New Horizons launched in 2006 and crossed Neptune's orbit in 2014. Although they have not entered interstellar space yet, Pioneer 10, Pioneer 11, and New Horizons are heading toward this landmark. Significant encounters between the DSN transmissions and nearby stars are highlighted below, and the full list of encounters for each spacecraft is available in the electronic version of this paper.

Signals from Earth constantly leak into space. For example, 5 Megawatt UHF television picture signals have an effective radiated power of 5 x $10^{6}$ W and an effective isotropic radiated power (EIRP) of approximately 8 x $10^{6}$ W, while DSN transmissions at 20 kW power have an EIRP of $~10^{10}$ W (\cite{Enriquez2017}, Equation 6). Outgoing signals from Earth to the interstellar spacecraft have an EIRP approximately $10^{3}$ times stronger than that of typical leakage, so they have a much higher chance of being noticed by intelligent extraterrestrial life \citep{Sullivan1978}.

Inspired by the work of \cite{Bailer-Jones2019} and their calculations of the future trajectories of our interstellar spacecraft, we use similar principles to identify a set of SETI targets that are in the background of sky positions occupied by NASA spacecraft when transmissions via the DSN were ongoing. Previous SETI target papers have identified stellar populations that are of particular interest. \cite{Sheikh2020} identify stars in the Earth-Transit Zone that would potentially know of the presence of Earth by observing the Earth transit the Sun. Similarly, \cite{Suphapolthaworn2022} identify stars that may see the microlens signal of Earth. A corollary of the Earth-Transit Zone theory has lead to the SETI observations of transiting planets \citep{Traas2021,Siemion2013}, with their ecliptic planes aligned with our line of sight. As the majority of the Earth's outward radio transmissions are near the ecliptic, such as communication with interplanetary spacecraft, perhaps other civilizations are sending radio transmissions in a similar way. Proximity is also an important property when selecting SETI targets \citep{Isaacson2017} due to the diminishing power of radio transmissions with the square of the distance to the target.

\section{Methods}
\subsection{Calculating Beamwidth} \label{sec:beamwidth}

We compared our beamwidth calculations with the referenced value from DSN and have determined to use a beamwidth of 0.128° $\pm$ 0.013° in our cone search around the ephemerides of each spacecraft. The beamwidth is the angle of a transmission at which the majority of the transmission’s power radiates. By calculating beamwidth, we can identify the angle on the sky over which the transmissions from Earth to the spacecraft are spread. This angle is used to search circular areas on the sky centered around the spacecraft using half of the beamwidth as the radius.

To calculate the half-power beamwidth of transmissions, we use the equation $\theta = k\lambda/d$, where $\theta$ is the beamwidth, $k$ is a constant that depends on a parabolic antenna’s reflector shape, $\lambda$ is wavelength, and $d$ is the diameter of the parabolic antenna. A parabolic antenna’s $k$ constant is typically 70, which is the value we use in our calculations. The uplink transmissions from Earth to Voyager 1 have a frequency of 2114.677 MHz\footnote{https://descanso.jpl.nasa.gov/DPSummary/Descanso4--Voyager\_ed.pdf}, so we use a wavelength of 0.142 m. The uplink transmissions to the other spacecraft fall within 4 MHz of this number, minimally impacting the beam width. The diameter of all of the DSN dishes used to communicate with the various spacecraft is 70 m. Using these values to manually calculate the beamwidth of the transmissions, we obtained a beamwidth angle of 0.142 degrees.

According to a DSN report on its 70 m subnet telecommunications interfaces\footnote{https://deepspace.jpl.nasa.gov/dsndocs/810-005/101/101E.pdf}, the half-power beamwidth on the S-band for transmissions from its 70 meter diameter dishes is 0.128° $\pm$ 0.013°. This value supports our manual calculation, just beyond one-sigma agreement. Assuming that DSN uses more technical procedures to calculate the half-power beamwidth, we adopt DSN’s value of 0.128° as the beamwidth angle during our search. 

\subsection{Cone Search Construction}

We use the JPL Horizons System to produce ephemerides for each spacecraft through their start and end dates of communication with the DSN. We then perform a cone search around each position in the ephemerides using the Gaia data query to find stars that will encounter DSN transmissions. To ensure our search covered the entire area behind each spacecraft, we used a stepsize equal to the beamwidth of 460\arcsec to create the ephemerides.

For each spacecraft, the ephemeris begins on that spacecraft’s launch date and ends on the last date of communication or the last date provided by the JPL Horizons System. Although DSN did not immediately begin communicating with the spacecraft upon launch, as it focuses on objects beyond Earth's orbit, we assume that other similar transmitters sent signals to the spacecraft during the earlier parts of their journeys. The DSN was not in constant communication with the spacecraft, but we assume constant communication in our search. This should not greatly impact results due to the large area on the sky covered by a single transmission at the distance to the average star. Table \ref{tab:date_table} displays the start and end dates of the ephemerides. 

The ephemerides for Voyager 1 and 2 end on the last date provided by the JPL Horizons System, although NASA predicts that both spacecraft will be in range of DSN for 6 additional years\footnote{https://voyager.jpl.nasa.gov/frequently-asked-questions/}. However, it is unlikely many new stars will be encountered by the transmissions during this time as the positions of the spacecraft on the sky will be nearly constant. It should also be noted that there is a gap in the ephemeris for Voyager 2 from 2020 Mar 9 to 2020 Oct 29\footnote{https://www.nasa.gov/feature/jpl/nasa-contacts-voyager-2-using-upgraded-deep-space-network-dish}. Earth did not send transmissions to Voyager 2 during this time because the dish used to communicate with Voyager 2 was being upgraded. 

The ephemerides for Pioneer 10 and 11 end on Earth’s final day of communication with the spacecraft\footnote{https://nssdc.gsfc.nasa.gov/nmc/spacecraft/display.action?id=1972-012A}\footnote{https://solarsystem.nasa.gov/missions/pioneer-11/in-depth}. The ephemeris for New Horizons ends on the last searchable date provided by JPL. It is predicted that the New Horizons project will continue through the late 2030s\footnote{https://www.nasa.gov/feature/nasa-s-new-horizons-reaches-a-rare-space-milestone/}, so we note that more stars will be encountered by the transmissions from Earth to New Horizons beyond our search. 

In this search we queried the GCNS, a catalogue under the European Space Agency’s Gaia mission that includes approximately 300,000 precisely mapped stars within 100 pc of Earth. We filtered our search to only return stars with a parallax less than 10 milliarcseconds to ensure identified stars fell within 100 pc. Stars with a parallax error greater than 0.34 mas were also excluded, leaving approximately 35,000 stars in our search \cite{Bailer-Jones2019}. This filter was applied only to the stars in the results of our search \citep{Bailer-Jones2019}. Applying this filter based on parallax error decreased the number of stars encountered by transmissions to Voyager 1 by 21, to Voyager 2 by 26, to Pioneer 10 by 25, to Pioneer 11 by 26, and to New Horizons by 9. 

\begin{table}[ht]
    \centering
\begin{tabular}{clc}
 \hline
 \multicolumn{3}{c}{Ephemerides' Start and End Dates} \\
 \hline
 \hline
 Spacecraft   & Start Date & End Date\\ \hline
 Voyager 1    & 1977 Sep 6 & 2030 Dec 31\\
 Voyager 2    & 1977 Aug 21 & 2030 Dec 31\\
 Pioneer 10   & 1972 Apr 3 & 2003 Jan 22\\
 Pioneer 11   & 1973 Apr 7 & 1995 Sep 30\\
 New Horizons & 2006 Jan 20 & 2024 Dec 31\\ \hline
\end{tabular} 
\caption{The start and end dates of the ephemerides for each spacecraft in our search.}
\label{tab:date_table}
\end{table}

\begin{figure}[ht]
\centering
\includegraphics[scale=0.6]{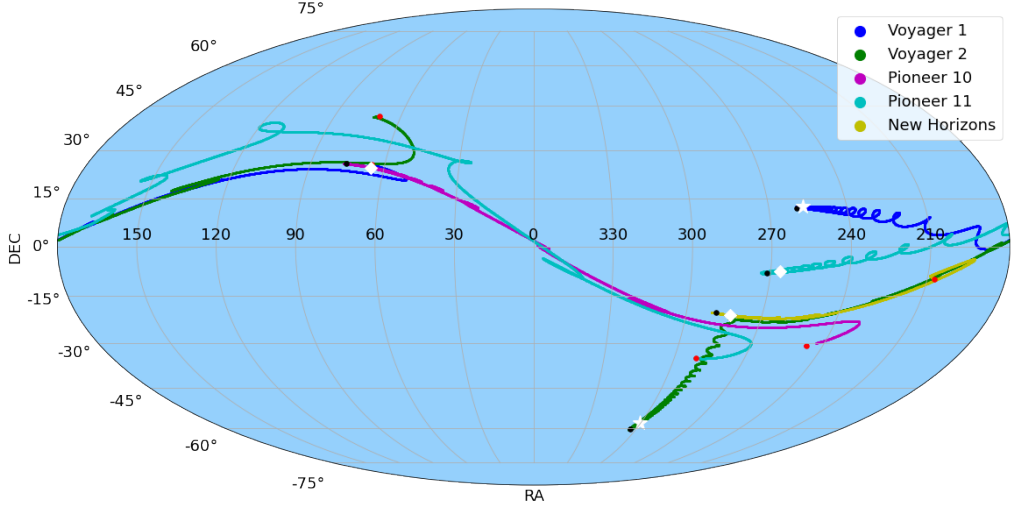}
\caption{The ephemerides of all five spacecraft on the Mollweide projection from the beginning to the end of the dates used for this search. Red points highlight the start dates of each ephemeris. Black points highlight end dates. White stars highlight the dates when Voyager 1 and Voyager 2 leave the heliosphere. White diamonds highlight the dates when Pioneer 10, Pioneer 11, and New Horizons crossed Neptune's orbit. The legend shows which color correlates to which spacecraft's ephemeris.}
\label{fig:combined_ephemeris}
\end{figure}

Figure \ref{fig:combined_ephemeris} shows the ephemerides for all the spacecraft on the Mollweide projection, highlighting the start and end dates of all the spacecraft in our search.  The dates that Voyager 1 and 2 entered interstellar space are 2012 Aug 1 and 2018 Dec 10, respectively. The dates when Pioneer 10, Pioneer 11, and New Horizons crossed Neptune's orbit are 1983 Jun 13, 1990 Feb 23, and 2014 Aug 25, respectively. All five spacecraft trace out at least a portion of the ecliptic plane during portions of their journeys. Each spacecraft’s change in position becomes very small toward the end of its ephemeris.

\subsection{Performing the Cone Search} 
\label{sec:search}

We perform our cone search using Astronomical Data Query Language (ADQL) and advance the positions of the stars to obtain the stars encountered by transmissions from DSN to the spacecraft. 

We first ran our cone search with a radius of 0.453 degrees, or 1630 arcseconds. This radius was chosen based on Barnard's star, a star that has the highest proper motion at 10.3\arcsec per year. At the maximum distance of 100 pc that our search includes, Barnard's star would travel 3260\arcsec before transmissions encountered it. The selected radius is half of this value. This search returned stars that fell within a circle centered at an ephemeris point with the given radius. We iterated this search over each position of the spacecrafts' ephemerides in order to search the entire area behind the spacecraft. 

We next used Gaia proper motions to advance the position of each star to where it would be at the time of encounter. If this position was further than 0.064° (half of the transmissions' beamwidth) from the center of the cone search, it was removed from our results. This guarantees that all of the stars in our results are within the projected area of the transmissions at the time of encounter.

\section{Results and Discussion}

\begin{figure}[ht]
\centering
\includegraphics[scale=0.6]{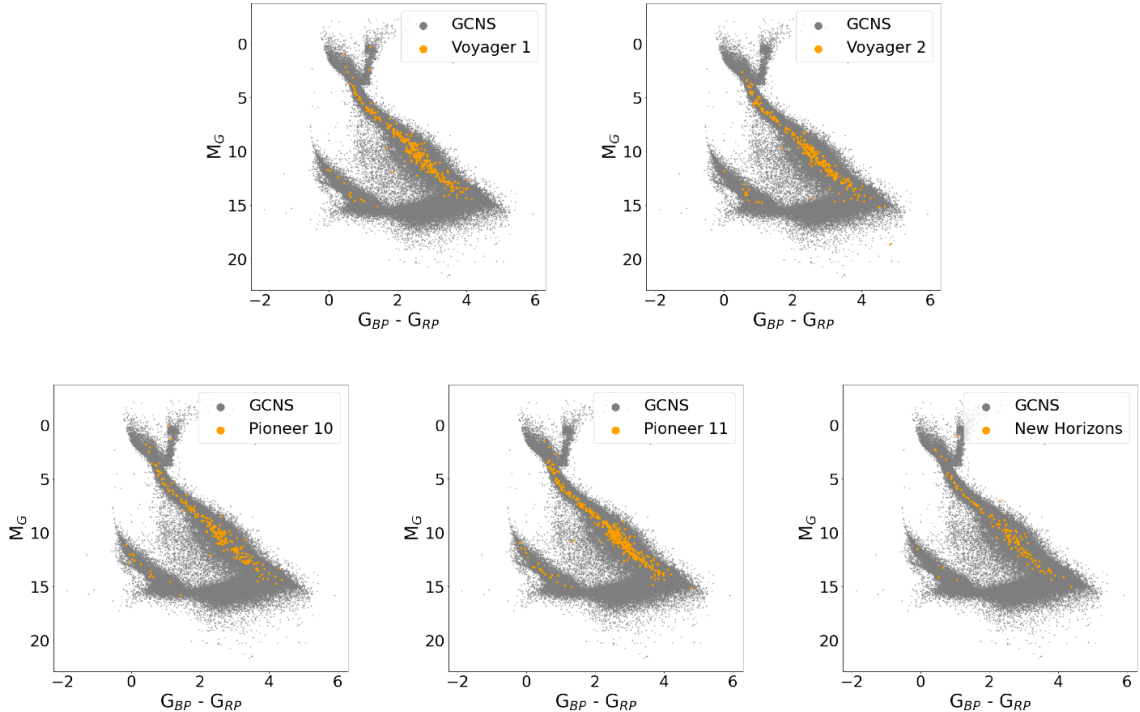}
    \caption{Color-magnitude diagrams for the stars encountered by each spacecraft. The color of the stars, red magnitude subtracted from blue magnitude, is graphed on the x-axis. The absolute magnitude of the stars is graphed on the y-axis. The grey points in the background represent the entire group of more than $\sim300,000$ stars within 100 pc in the GCNS. The orange points represent the stars specifically encountered by the transmissions of each spacecraft. See legend for which diagram correlates to each spacecraft.}
    \label{fig:superset_hr}
\end{figure}

We analyze the characteristics of the stars encountered by spacecraft transmissions through various diagrams and highlight encounters that have already occurred. 
Figure \ref{fig:superset_hr} shows a color-magnitude diagram for the group of stars encountered by each spacecraft. In the background of each diagram, the entirety of the GCNS is displayed. The GCNS catalogue includes 331,312 stars within 100pc. It is complete for stars brighter than M8 and contains 92\% of the M9 dwarfs within 100 pc of Earth's sun (\cite{gaia_gcns}). The color-magnitude diagrams for each spacecraft look similar in their distributions. Stars appear along the main sequence and within the white dwarf group. One noticeable difference among these diagrams is the number of stars encountered. Pioneer 11 has the most populous group, encountering 411 stars, and New Horizons has the least populous group, encountering 142 stars. Voyager 1, Voyager 2, and Pioneer 10 will encounter a total of 289, 325, and 241 stars, respectively. The five stars encountered earliest by each spacecraft, these stars' distances, and the contact and return dates of the spacecraft transmissions are summarized in Table \ref{star_table}.

%

\begin{figure}[ht]
\centering
\includegraphics[scale=0.6]{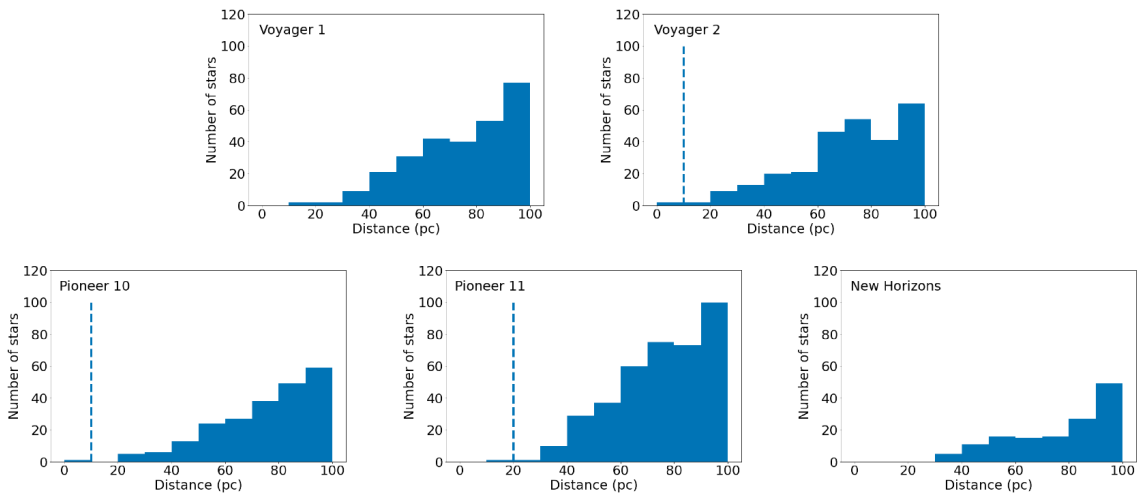}
    \caption{Distance histograms for the stars encountered by each spacecraft, ranging from 0 to 100 pc. Stars left of the blue dashed lines have already been encountered by the spacecraft's transmissions. The transmissions of Voyager 1 and New Horizons have not yet encountered stars. The transmissions of Voyager 2, Pioneer 10, and Pioneer 11 have already encountered 2 stars within 10 pc, 1 star within 10 pc, and 1 star within 20 pc, respectively. See labels on the top left corners for which diagram correlates to each spacecraft's encountered stars.}
    \label{fig:dist_hist}
\end{figure}

Figure \ref{fig:dist_hist} displays distance histograms for the stars encountered by each spacecraft's transmissions. Each histogram has bins of 10 pc in the range 0 to 100 pc. The distributions of stars on the diagrams for each spacecraft are similar and reflect the stellar population overall. The diagrams include far fewer stars with smaller distances. As distance increases, the encompassed number of stars increases significantly, due to the ever increasing volume as a function of distance. Only Voyager 2 and Pioneer 10 transmissions encounter any stars within 10 pc. 

%

\subsection{Significant Encounters}
\label{sig_encounters}
We highlight the first stars that will be reached by the transmissions of each spacecraft, including some stars that have already been encountered. An overview of the number of stars that will be contacted and dates of contact is also provided. The transmissions to Voyager 2, Pioneer 10, and Pioneer 11 have already encountered at least one star, while those to Voyager 1 and New Horizons will encounter their first stars in the near future.

To calculate the date that each star will be reached by the transmissions, we consider that the speed of light covers one parsec every 3.26 years. This allowed us to determine the time in years that it will take the transmissions to reach each star. Adding this time to the date on which the transmission was sent to the respective spacecraft yields the date the star will encounter the transmission.

The transmissions of Voyager 1 have not yet reached any stars, but they will reach their first, Gaia EDR3 3814369840081992448 (an M-dwarf), in 2044 (Table \ref{star_table}). Voyager 1's transmissions will contact 277 stars by the year 2341. The earliest we can expect to receive a returned transmission from potential intelligent extraterrestrial life encountered by Voyager 1's transmissions is 2109.

The transmissions of Voyager 2 have already encountered two objects, Gaia EDR3 3698534434669937024 (an M-dwarf) and Gaia EDR3 6306068659857135232 (a brown dwarf). Both of these objects were reached in 2007. Voyager 2's transmissions will encounter 272 stars by 2336. The earliest we can expect to receive a returned transmission is 2033. 

The transmissions of Pioneer 10 encountered one white dwarf star, Gaia EDR3 2611561706216413696, in 2002. Pioneer 10's transmissions will encounter 222 stars by 2313. The earliest we can expect to receive a returned transmission is 2029.
.
The transmissions of Pioneer 11 also encountered one M-dwarf star, Gaia EDR3 640862653425455360, in 2018. Pioneer 11's transmissions will reach 386 stars by 2317. The earliest we can expect to receive a returned transmission is 2058.

New Horizons's transmissions have not yet encountered a star but will contact the M-dwarf Gaia EDR3 3618803417701443072 in 2119. New Horizons's transmissions will encounter 139 stars by 2338. The earliest we can expect to receive a returned transmission is 2232. Transmissions to New Horizons will not encounter a star until 75 years after the transmissions of any other spacecraft. This difference is primarily due to New Horizons having a later launch date and spending less time within the heliosphere. A larger number of stars are contacted while the spacecraft are within the heliosphere because this portion of their journeys covers a larger portion of the sky. New Horizons was launched in 2006, while the other spacecraft were launched in the 1970s. It took New Horizons eight years to cross Neptune's orbit, whereas it took Pioneer 10 and Pioneer 11 eleven and seventeen years, respectively. 

Table \ref{star_table} shows 25 stars encountered by the transmissions to the spacecraft, sorted by the year that we would expect a returned transmission. Only the five stars that will first be encountered by each of the spacecraft's transmissions are included. Shown are the identifiers of these stars, their distance from Earth, the year that they will be or were contacted, the year we would expect a returned transmission, the time the star spends in the projected area of the transmission beam, and the spectral type of the star. Full CSV tables containing all the stars contacted by transmissions to each spacecraft are available in the electronic version of this paper. The calculations are available on github. Active SETI surveys should consider observations of the contacted stars near dates listed in the table.

\begin{table}[ht]
    \centering

\begin{tabular}{cclcccc}
 \hline
 \multicolumn{7}{c}{Significant Star Encounters} \\
 \hline
 \hline
 Spacecraft & Gaia EDR3 & Distance & Contact & Return & Time Spent in  & Spectral\\
  & Source ID & (pc) & Year & Year & Beam (days) & Type\\
 \hline
 Pioneer 10 & 2611561706216413696 & 8.54 & 2002 & 2029 & 3.54 & DZ13\\
 Voyager 2 & 6306068659857135232 & 7.41 & 2007 & 2031 & 49.10 & L5V\\
 Voyager 2 & 3698534434669937024 & 8.09 & 2007 & 2033 & 11.92 & M4.5Ve\\
 Pioneer 11 & 640862653425455360 & 12.34 & 2018 & 2058 & 23.44 & M4\\
 Voyager 2 & 117709729140217216 & 15.35 & 2028 & 2078 & 5.01 &	M4V\\
 Voyager 1 & 3814369840081992448 & 19.95 & 2044 & 2109 & 27.68 & M3.5V\\
 Voyager 2 & 3429699684156117760 & 22.26 & 2050 & 2123 & 2.67 & M3\\
 Pioneer 10 & 148133112801536000 & 21.16 & 2055 & 2124 & 408.53 & M3.5V\\
 Pioneer 11 & 129200553364168064 & 22.93 & 2050 & 2125 & 3.05 & ---\\
 Voyager 1 & 4540984195244025088 & 18.41 & 2071 & 2131 & 5633.86 & M3.97\\
 Voyager 2 & 663767443044739840 & 23.49 & 2055 & 2132 & 7.70 & M4V\\
 Pioneer 10 & 6233452028580863744 & 24.88 & 2053 & 2134 & 2.91 & ---\\
 Pioneer 10 & 4110228589254067328 & 26.58 & 2059 & 2146 & 2.66 & ---\\
 Voyager 1 & 865699110037321600 & 25.84 & 2062 & 2147 & 3.16 & M1V\\
 Pioneer 10 & 84210874016975360 & 28.51 & 2070 & 2163 & 9.88 & M\\
 Voyager 1 & 3367558512468755968 & 30.23 & 2077 & 2175 & 3.00 & M4.5\\
 Voyager 1 & 1174143182830505984 & 30.04 & 2081 & 2179 & 20.43 & K0V\\
 Pioneer 11 & 3691265460219482240 & 32.84 & 2087 & 2195 & 7.13 & M5V\\
 Pioneer 11 & 100380639907919872 & 34.38 & 2087 & 2199 & 6.46 & G5\\
 Pioneer 11 & 2631967439437024384 & 36.30 & 2092 & 2210 & 7.05 & DC9\\
 New Horizons & 3618803417701443072 & 34.79 & 2119 & 2232 & 0.00025 & ---\\
 New Horizons & 4117657650084825472 & 36.41 & 2127 & 2246 & 0.00062 & ---\\
 New Horizons & 3635672468691948672 & 38.17 & 2130 & 2255 & 0.000036 & ---\\
 New Horizons & 4127569403769839488 & 38.12 & 2132 & 2256 & 0.00023 & ---\\
 New Horizons & 4084898525733384832 & 37.51 & 2135 & 2257 & 0.017 & ---\\ \hline
\end{tabular}

\caption{Stars encountered by the transmissions to the spacecraft, sorted by the year that we would expect a returned transmission. Only the 5 stars encountered earliest by each spacecraft are included in this list. Gaia EDR3 source ID, the distance of the star, the contact and return year of the transmission, the time spent by the star within the transmission's beam, and the spectral type of the star are reported. A CSV including all of our results is available in the electronic version of this paper.}
\label{star_table}
\end{table}

\begin{figure}[ht]
\centering
\includegraphics[scale=0.5]{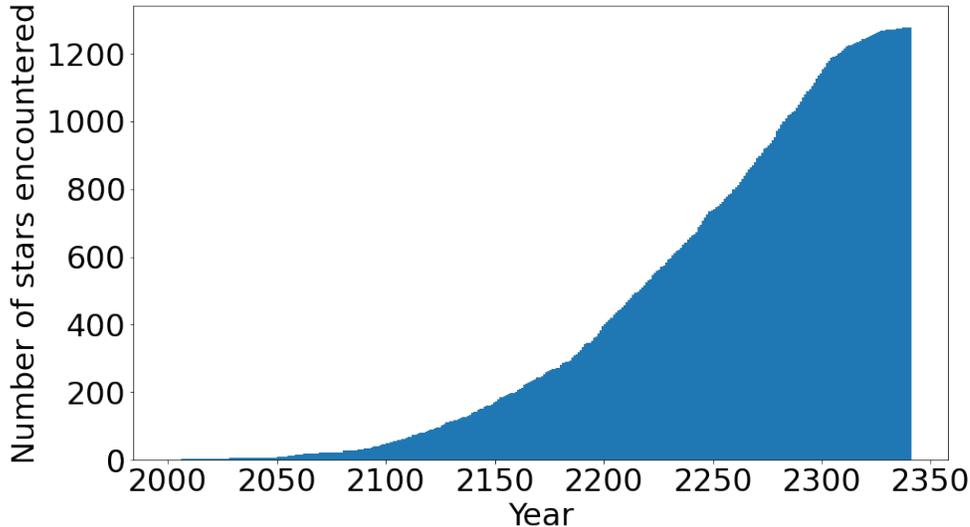}
\caption{The cumulative number of stars encountered by transmissions from DSN to the spacecraft as a function of time. The x-axis plots the year of encounter. As of January 2023, only 4 stars have been encountered by DSN transmissions.}
\label{fig:encountered_time}
\end{figure}

Figure \ref{fig:encountered_time} shows a cumulative histogram of the total number of stars encountered as a function of time. The y-axis plots the cumulative number of stars encountered by transmissions from DSN to the spacecraft. The total number of stars increases at a very low rate through 2100 and then at an increasingly greater rate. This shows that the majority of the encountered stars are at relatively large distances from Earth as it takes DSN transmissions longer to encounter them. This increase in stars contacted over time is consistent with the increasing volume covered by the transmissions over time.

\section{Conclusion} \label{sec:conclusion}

We are confident that the surrounding planets of the encountered stars will also encounter the spacecrafts' transmissions. The closest star encountered by the transmissions of the spacecraft, Gaia EDR3 6306068659857135232, is 7.41 pc away from Earth. At this distance, the 0.128° beam will cover a radius of 0.00827 pc or 1707.332 AU around the star. For context, Pluto is only 39 AU from our sun. As the beam travels farther to other stars, this radius will only grow, showing that we can assume that all of the planets orbiting each star will also encounter the spacecrafts' transmissions.

We compare the intensity of the DSN transmissions at 20 kW power, or $~10^{10}$ W EIRP (\cite{Enriquez2017}, Equation 6) to the planetary radar at Arecibo with an EIRP of $~10^{12}$ W. While the transmissions are a few orders of magnitude smaller in EIRP, the documented positions of the spacecraft under consideration reveal more details about the background stars for DSN transmission than the Arecibo planetary radar.

In the prioritization of stars in the sky to search for technosignatures, we provide a list of targets that have or will receive transmissions from the DSN that were targeted at our own interplanetary spacecraft. Transmissions to Voyager have already encountered an M-dwarf, GJ 1154, and a brown dwarf, Gaia EDR3 6306068659857135232. Transmissions to Pioneer 10 have encountered a white dwarf, GJ 1276. Transmissions to Pioneer 11 have encountered a M-dwarf, GJ 359. We have shown that the radio transmissions using the DSN are stronger than typical leakage and are useful for identifying good technosignature targets. Just as the future trajectories of the Voyager and Pioneer spacecraft have been calculated and their future interactions with distant stars cataloged by \cite{Bailer-Jones2019}, we now also consider the paths of DSN communications with those spacecraft to the stars beyond them.

\software{astropy \citep{2022ApJ...935..167A}, numpy \citep{harris2020array}, matplotlib \citep{Hunter:2007}, pandas \citep{mckinney2010data}, \url{https://github.com/reillyderrick/Transmission-Encounters} }

\section{Acknowledgement} \label{sec:acknowledgement}

 We thank the anonymous referee who provided useful comments that improved the quality of the paper. This research used the Gaia Database: {\it Gaia} (\url{https://www.cosmos.esa.int/gaia}), processed by the {\it Gaia} Data Processing and Analysis Consortium (DPAC, \url{https://www.cosmos.esa.int/web/gaia/dpac/consortium}). Funding for the DPAC has been provided by national institutions, in particular the institutions participating in the {\it Gaia} Multilateral Agreement. We want to thank Vishal Gajjar for helpful comments on the paper. RD developed the computer code, created the graphics and wrote the text of the paper. HI provided scientific guidance and editing of the manuscript. HI is supported by Breakthrough Listen which is managed by the Breakthrough Initiatives, sponsored by the Breakthrough Prize Foundation. 

\bibliography{main}{}

\begin{thebibliography}{}
\expandafter\ifx\csname natexlab\endcsname\relax\def\natexlab#1{#1}\fi
\providecommand{\url}[1]{\href{#1}{#1}}
\providecommand{\dodoi}[1]{doi:~\href{http://doi.org/#1}{\nolinkurl{#1}}}
\providecommand{\doeprint}[1]{\href{http://ascl.net/#1}{\nolinkurl{http://ascl.net/#1}}}
\providecommand{\doarXiv}[1]{\href{https://arxiv.org/abs/#1}{\nolinkurl{https://arxiv.org/abs/#1}}}

\bibitem[{{Astropy Collaboration} {et~al.}(2022){Astropy Collaboration},
  {Price-Whelan}, {Lim}, {Earl}, {Starkman}, {Bradley}, {Shupe}, {Patil},
  {Corrales}, {Brasseur}, {N{\"o}the}, {Donath}, {Tollerud}, {Morris},
  {Ginsburg}, {Vaher}, {Weaver}, {Tocknell}, {Jamieson}, {van Kerkwijk},
  {Robitaille}, {Merry}, {Bachetti}, {G{\"u}nther}, {Aldcroft},
  {Alvarado-Montes}, {Archibald}, {B{\'o}di}, {Bapat}, {Barentsen},
  {Baz{\'a}n}, {Biswas}, {Boquien}, {Burke}, {Cara}, {Cara}, {Conroy},
  {Conseil}, {Craig}, {Cross}, {Cruz}, {D'Eugenio}, {Dencheva}, {Devillepoix},
  {Dietrich}, {Eigenbrot}, {Erben}, {Ferreira}, {Foreman-Mackey}, {Fox},
  {Freij}, {Garg}, {Geda}, {Glattly}, {Gondhalekar}, {Gordon}, {Grant},
  {Greenfield}, {Groener}, {Guest}, {Gurovich}, {Handberg}, {Hart},
  {Hatfield-Dodds}, {Homeier}, {Hosseinzadeh}, {Jenness}, {Jones}, {Joseph},
  {Kalmbach}, {Karamehmetoglu}, {Ka{\l}uszy{\'n}ski}, {Kelley}, {Kern},
  {Kerzendorf}, {Koch}, {Kulumani}, {Lee}, {Ly}, {Ma}, {MacBride}, {Maljaars},
  {Muna}, {Murphy}, {Norman}, {O'Steen}, {Oman}, {Pacifici}, {Pascual},
  {Pascual-Granado}, {Patil}, {Perren}, {Pickering}, {Rastogi}, {Roulston},
  {Ryan}, {Rykoff}, {Sabater}, {Sakurikar}, {Salgado}, {Sanghi}, {Saunders},
  {Savchenko}, {Schwardt}, {Seifert-Eckert}, {Shih}, {Jain}, {Shukla}, {Sick},
  {Simpson}, {Singanamalla}, {Singer}, {Singhal}, {Sinha}, {Sip{\H{o}}cz},
  {Spitler}, {Stansby}, {Streicher}, {{\v{S}}umak}, {Swinbank}, {Taranu},
  {Tewary}, {Tremblay}, {Val-Borro}, {Van Kooten}, {Vasovi{\'c}}, {Verma}, {de
  Miranda Cardoso}, {Williams}, {Wilson}, {Winkel}, {Wood-Vasey}, {Xue},
  {Yoachim}, {Zhang}, {Zonca}, \& {Astropy Project
  Contributors}}]{2022ApJ...935..167A}
{Astropy Collaboration}, {Price-Whelan}, A.~M., {Lim}, P.~L., {et~al.} 2022,
  \apj, 935, 167, \dodoi{10.3847/1538-4357/ac7c74}

\bibitem[{{Bailer-Jones} \& {Farnocchia}(2019)}]{Bailer-Jones2019}
{Bailer-Jones}, C. A.~L., \& {Farnocchia}, D. 2019, Research Notes of the
  American Astronomical Society, 3, 59, \dodoi{10.3847/2515-5172/ab158e}

\bibitem[{{Enriquez} {et~al.}(2017){Enriquez}, {Siemion}, {Foster}, {Gajjar},
  {Hellbourg}, {Hickish}, {Isaacson}, {Price}, {Croft}, {DeBoer}, {Lebofsky},
  {MacMahon}, \& {Werthimer}}]{Enriquez2017}
{Enriquez}, J.~E., {Siemion}, A., {Foster}, G., {et~al.} 2017, \apj, 849, 104,
  \dodoi{10.3847/1538-4357/aa8d1b}

\bibitem[{{Gaia Collaboration} {et~al.}(2021){Gaia Collaboration}, {Smart},
  {Sarro}, {Rybizki}, {Reyl{\'e}}, {Robin}, {Hambly}, {Abbas}, {Barstow}, {de
  Bruijne}, {Bucciarelli}, {Carrasco}, {Cooper}, {Hodgkin}, {Masana},
  {Michalik}, {Sahlmann}, {Sozzetti}, {Brown}, {Vallenari}, {Prusti},
  {Babusiaux}, {Biermann}, {Creevey}, {Evans}, {Eyer}, {Hutton}, {Jansen},
  {Jordi}, {Klioner}, {Lammers}, {Lindegren}, {Luri}, {Mignard}, {Panem},
  {Pourbaix}, {Randich}, {Sartoretti}, {Soubiran}, {Walton}, {Arenou},
  {Bailer-Jones}, {Bastian}, {Cropper}, {Drimmel}, {Katz}, {Lattanzi}, {van
  Leeuwen}, {Bakker}, {Casta{\~n}eda}, {De Angeli}, {Ducourant}, {Fabricius},
  {Fouesneau}, {Fr{\'e}mat}, {Guerra}, {Guerrier}, {Guiraud}, {Jean-Antoine
  Piccolo}, {Messineo}, {Mowlavi}, {Nicolas}, {Nienartowicz}, {Pailler},
  {Panuzzo}, {Riclet}, {Roux}, {Seabroke}, {Sordo}, {Tanga}, {Th{\'e}venin},
  {Gracia-Abril}, {Portell}, {Teyssier}, {Altmann}, {Andrae}, {Bellas-Velidis},
  {Benson}, {Berthier}, {Blomme}, {Brugaletta}, {Burgess}, {Busso}, {Carry},
  {Cellino}, {Cheek}, {Clementini}, {Damerdji}, {Davidson}, {Delchambre},
  {Dell'Oro}, {Fern{\'a}ndez-Hern{\'a}ndez}, {Galluccio}, {Garc{\'\i}a-Lario},
  {Garcia-Reinaldos}, {Gonz{\'a}lez-N{\'u}{\~n}ez}, {Gosset}, {Haigron},
  {Halbwachs}, {Harrison}, {Hatzidimitriou}, {Heiter}, {Hern{\'a}ndez},
  {Hestroffer}, {Holl}, {Jan{\ss}en}, {Jevardat de Fombelle}, {Jordan},
  {Krone-Martins}, {Lanzafame}, {L{\"o}ffler}, {Lorca}, {Manteiga}, {Marchal},
  {Marrese}, {Moitinho}, {Mora}, {Muinonen}, {Osborne}, {Pancino}, {Pauwels},
  {Recio-Blanco}, {Richards}, {Riello}, {Rimoldini}, {Roegiers}, {Siopis},
  {Smith}, {Ulla}, {Utrilla}, {van Leeuwen}, {van Reeven}, {Abreu Aramburu},
  {Accart}, {Aerts}, {Aguado}, {Ajaj}, {Altavilla}, {{\'A}lvarez}, {{\'A}lvarez
  Cid-Fuentes}, {Alves}, {Anderson}, {Anglada Varela}, {Antoja}, {Audard},
  {Baines}, {Baker}, {Balaguer-N{\'u}{\~n}ez}, {Balbinot}, {Balog}, {Barache},
  {Barbato}, {Barros}, {Bartolom{\'e}}, {Bassilana}, {Bauchet},
  {Baudesson-Stella}, {Becciani}, {Bellazzini}, {Bernet}, {Bertone}, {Bianchi},
  {Blanco-Cuaresma}, {Boch}, {Bombrun}, {Bossini}, {Bouquillon}, {Bragaglia},
  {Bramante}, {Breedt}, {Bressan}, {Brouillet}, {Burlacu}, {Busonero},
  {Butkevich}, {Buzzi}, {Caffau}, {Cancelliere}, {C{\'a}novas},
  {Cantat-Gaudin}, {Carballo}, {Carlucci}, {Carnerero}, {Casamiquela},
  {Castellani}, {Castro-Ginard}, {Castro Sampol}, {Chaoul}, {Charlot},
  {Chemin}, {Chiavassa}, {Cioni}, {Comoretto}, {Cornez}, {Cowell}, {Crifo},
  {Crosta}, {Crowley}, {Dafonte}, {Dapergolas}, {David}, {David}, {de Laverny},
  {De Luise}, {De March}, {De Ridder}, {de Souza}, {de Teodoro}, {de Torres},
  {del Peloso}, {del Pozo}, {Delgado}, {Delgado}, {Delisle}, {Di Matteo},
  {Diakite}, {Diener}, {Distefano}, {Dolding}, {Eappachen}, {Edvardsson},
  {Enke}, {Esquej}, {Fabre}, {Fabrizio}, {Faigler}, {Fedorets}, {Fernique},
  {Fienga}, {Figueras}, {Fouron}, {Fragkoudi}, {Fraile}, {Franke}, {Gai},
  {Garabato}, {Garcia-Gutierrez}, {Garc{\'\i}a-Torres}, {Garofalo}, {Gavras},
  {Gerlach}, {Geyer}, {Giacobbe}, {Gilmore}, {Girona}, {Giuffrida}, {Gomel},
  {Gomez}, {Gonzalez-Santamaria}, {Gonz{\'a}lez-Vidal}, {Granvik},
  {Guti{\'e}rrez-S{\'a}nchez}, {Guy}, {Hauser}, {Haywood}, {Helmi}, {Hidalgo},
  {Hilger}, {H{\l}adczuk}, {Hobbs}, {Holland}, {Huckle}, {Jasniewicz},
  {Jonker}, {Juaristi Campillo}, {Julbe}, {Karbevska}, {Kervella}, {Khanna},
  {Kochoska}, {Kontizas}, {Kordopatis}, {Korn}, {Kostrzewa-Rutkowska},
  {Kruszy{\'n}ska}, {Lambert}, {Lanza}, {Lasne}, {Le Campion}, {Le Fustec},
  {Lebreton}, {Lebzelter}, {Leccia}, {Leclerc}, {Lecoeur-Taibi}, {Liao},
  {Licata}, {Lindstr{\o}m}, {Lister}, {Livanou}, {Lobel}, {Madrero Pardo},
  {Managau}, {Mann}, {Marchant}, {Marconi}, {Marcos Santos}, {Marinoni},
  {Marocco}, {Marshall}, {Martin Polo}, {Mart{\'\i}n-Fleitas}, {Masip},
  {Massari}, {Mastrobuono-Battisti}, {Mazeh}, {McMillan}, {Messina}, {Millar},
  {Mints}, {Molina}, {Molinaro}, {Moln{\'a}r}, {Montegriffo}, {Mor},
  {Morbidelli}, {Morel}, {Morris}, {Mulone}, {Munoz}, {Muraveva}, {Murphy},
  {Musella}, {Noval}, {Ord{\'e}novic}, {Orr{\`u}}, {Osinde}, {Pagani},
  {Pagano}, {Palaversa}, {Palicio}, {Panahi}, {Pawlak}, {Pe{\~n}alosa
  Esteller}, {Penttil{\"a}}, {Piersimoni}, {Pineau}, {Plachy}, {Plum},
  {Poggio}, {Poretti}, {Poujoulet}, {Pr{\v{s}}a}, {Pulone}, {Racero},
  {Ragaini}, {Rainer}, {Raiteri}, {Rambaux}, {Ramos}, {Ramos-Lerate}, {Re
  Fiorentin}, {Regibo}, {Ripepi}, {Riva}, {Rixon}, {Robichon}, {Robin},
  {Roelens}, {Rohrbasser}, {Romero-G{\'o}mez}, {Rowell}, {Royer}, {Rybicki},
  {Sadowski}, {Sagrist{\`a} Sell{\'e}s}, {Salgado}, {Salguero}, {Samaras},
  {Sanchez Gimenez}, {Sanna}, {Santove{\~n}a}, {Sarasso}, {Schultheis},
  {Sciacca}, {Segol}, {Segovia}, {S{\'e}gransan}, {Semeux}, {Shahaf},
  {Siddiqui}, {Siebert}, {Siltala}, {Slezak}, {Solano}, {Solitro}, {Souami},
  {Souchay}, {Spagna}, {Spoto}, {Steele}, {Steidelm{\"u}ller}, {Stephenson},
  {S{\"u}veges}, {Szabados}, {Szegedi-Elek}, {Taris}, {Tauran}, {Taylor},
  {Teixeira}, {Thuillot}, {Tonello}, {Torra}, {Torra}, {Turon}, {Unger},
  {Vaillant}, {van Dillen}, {Vanel}, {Vecchiato}, {Viala}, {Vicente},
  {Voutsinas}, {Weiler}, {Wevers}, {Wyrzykowski}, {Yoldas}, {Yvard}, {Zhao},
  {Zorec}, {Zucker}, {Zurbach}, \& {Zwitter}}]{gaia_gcns}
{Gaia Collaboration}, {Smart}, R.~L., {Sarro}, L.~M., {et~al.} 2021, \aap, 649,
  A6, \dodoi{10.1051/0004-6361/202039498}

\bibitem[{Harris {et~al.}(2020)Harris, Millman, van~der Walt, Gommers,
  Virtanen, Cournapeau, Wieser, Taylor, Berg, Smith, Kern, Picus, Hoyer, van
  Kerkwijk, Brett, Haldane, del R{\'{i}}o, Wiebe, Peterson,
  G{\'{e}}rard-Marchant, Sheppard, Reddy, Weckesser, Abbasi, Gohlke, \&
  Oliphant}]{harris2020array}
Harris, C.~R., Millman, K.~J., van~der Walt, S.~J., {et~al.} 2020, Nature, 585,
  357, \dodoi{10.1038/s41586-020-2649-2}

\bibitem[{Hunter(2007)}]{Hunter:2007}
Hunter, J.~D. 2007, Computing in Science \& Engineering, 9, 90,
  \dodoi{10.1109/MCSE.2007.55}

\bibitem[{{Isaacson} {et~al.}(2017){Isaacson}, {Siemion}, {Marcy}, {Lebofsky},
  {Price}, {MacMahon}, {Croft}, {DeBoer}, {Hickish}, {Werthimer}, {Sheikh},
  {Hellbourg}, \& {Enriquez}}]{Isaacson2017}
{Isaacson}, H., {Siemion}, A. P.~V., {Marcy}, G.~W., {et~al.} 2017, \pasp, 129,
  054501, \dodoi{10.1088/1538-3873/aa5800}

\bibitem[{McKinney {et~al.}(2010)}]{mckinney2010data}
McKinney, W., {et~al.} 2010, in Proceedings of the 9th Python in Science
  Conference, Vol. 445, Austin, TX, 51--56

\bibitem[{{Sheikh} {et~al.}(2020){Sheikh}, {Siemion}, {Enriquez}, {Price},
  {Isaacson}, {Lebofsky}, {Gajjar}, \& {Kalas}}]{Sheikh2020}
{Sheikh}, S.~Z., {Siemion}, A., {Enriquez}, J.~E., {et~al.} 2020, \aj, 160, 29,
  \dodoi{10.3847/1538-3881/ab9361}

\bibitem[{{Siemion} {et~al.}(2013){Siemion}, {Demorest}, {Korpela},
  {Maddalena}, {Werthimer}, {Cobb}, {Howard}, {Langston}, {Lebofsky}, {Marcy},
  \& {Tarter}}]{Siemion2013}
{Siemion}, A. P.~V., {Demorest}, P., {Korpela}, E., {et~al.} 2013, \apj, 767,
  94, \dodoi{10.1088/0004-637X/767/1/94}

\bibitem[{Sullivan {et~al.}(1978)Sullivan, Brown, \& Wetherill}]{Sullivan1978}
Sullivan, W.~T., Brown, S., \& Wetherill, C. 1978, Science, 199, 377.
\newblock \url{http://www.jstor.org/stable/1745785}

\bibitem[{{Suphapolthaworn} {et~al.}(2022){Suphapolthaworn}, {Awiphan},
  {Chatchadanoraset}, {Kerins}, {Specht}, {Nakharutai}, {Komonjinda}, \&
  {Robin}}]{Suphapolthaworn2022}
{Suphapolthaworn}, S., {Awiphan}, S., {Chatchadanoraset}, T., {et~al.} 2022,
  \mnras, 515, 5927, \dodoi{10.1093/mnras/stac1855}

\bibitem[{{Traas} {et~al.}(2021){Traas}, {Croft}, {Gajjar}, {Isaacson},
  {Lebofsky}, {MacMahon}, {Perez}, {Price}, {Sheikh}, {Siemion}, {Smith},
  {Drew}, \& {Worden}}]{Traas2021}
{Traas}, R., {Croft}, S., {Gajjar}, V., {et~al.} 2021, \aj, 161, 286,
  \dodoi{10.3847/1538-3881/abf649}

\end{thebibliography}
\bibliographystyle{aasjournal}

\end{document}